# Topological characteristics of local atomic arrangements at crystalline-amorphous structural transition in graphite

A.D. Rud*, I.M. Kirian, A.M. Lakhnik
*G.V. Kurdyumov Institute for Metal Physics of NASU,*
*36 academician Vernadsky blvd., 03680 Kiev, Ukraine*

**Abstract**

Quasi-continuous structural transformation from the crystalline to amorphous state takes place in graphite during ball-milling. The quantitative characteristics of a short- and a medium-range orders in carbon nanomaterials structure are determined by a combined application of X-ray diffraction analysis, reverse Monte Carlo modeling and Voronoi diagram method. High resolution TEM images revealed formation of globular carbon materials having onion-like structure. The Voronoi polyhedra (VP) constructed for simulated atomic configurations of the ball-milled graphite have an extraordinary variety in their topological and metric characteristics and contain a lot of 5-fold faces. The analysis of VP sphericity coefficient $K_{sph}$ enables a conclusion about the change of the local atomic arrangement in a structure of ball-milled carbon from graphite- to diamond-like. The sphericity coefficient is proposed to be as a parameter of the topological order to quantitative estimation of disordering degree in amorphous structures.

## 1. Introduction

Amorphous materials are an object of extensive investigation during the last decades due to their unique exploitative properties [1,2]. Mechanical properties of metallic glasses and bulk amorphous materials are about two to three times higher than those in crystalline states [2,3]. It is commonly recognized their properties are caused by the features of their atomic structures. Numerous researches were undertaken to clarify the atomic structure of disordered materials with the purpose to find the basic principles of amorphous structure formation [4-7]. The additional issue of the amorphous materials is their characterization of local atomic arrangements [2,8-13].

It was shown [14-16] that atomic arrangements of amorphous materials can be more adequately characterized by a medium-range order, which describes the most characteristic atomic configurations at a distance of 1-2 nm (several coordination spheres), compared to a short-range order (one or two coordination spheres). At the same time, the quantitative characteristics (an average coordination number and a bond length) for both short-range and medium-range orders cannot provide information about a local atomic arrangement in disordered materials and its topological characteristics. A combination of experimental techniques (high resolution electron microscopy, nanobeam electron diffraction analysis, atom probe tomography) and Voronoi-Delaunay analysis of modeling ensembles of atomic configurations reconstructed by reverse Monte Carlo (RMC) or molecular dynamics (MD) methods from experimental diffraction data is used for these objectives [2,12,17-23]. It should be also noted that these methods have been mainly applied for studies of multicomponent metallic glasses. In the other words, these materials have both chemical (composition) and topological short-range orders [2,4].

The order parameter $\eta_{AB}$ based on partial coordination numbers was proposed for the quantitative characterization of the chemical short-range order in amorphous systems [24]. This parameter is similar to the classical formula for the atomic short-range order in crystalline binary systems [25]. The problem of description of the topological short-range order in amorphous systems is significantly more complicated than that of chemical order, because selection of the order parameter is not obvious.

*Corresponding author:
E-mail address: rud@imp.kiev.ua (A. Rud)

Notice that conventional procedure for quantitative characterization of the topological order in amorphous materials allowing to distinction one amorphous structure from others is absent up to now. Taking into account the existence of interrelation between topological and chemical short-range orders, the investigation of the quantitative characteristics of the local atomic arrangement in amorphous uni-component system is advisable for resolving this problem.

Carbon based materials have a great interest through a broad spectrum of allotropic modifications depended on the degree of sp-hybridization of valence electrons [26]. Moreover, the amorphous carbon can be easily produced by either high-energy ball-milling (BM) of graphite [27-30] or high-voltage electric discharge treatment of hydrocarbon precursors [31-35]. The local atomic structure and topological characteristics of amorphous carbon produced by high-frequency electric discharge treatment of gaseous hydrocarbons [36] and high-temperature annealing of fullerenes $C_{60}$ [37,38] were investigated by X-ray, RMC and Voronoi-Delaunay methods. It was shown that there is a wide range of coordination polyhedra with predomination of pentahedral ones [36].

The goal of this paper is the systematic study of the evolution of local atomic arrangements during the quasi-continuous crystalline-amorphous transition under the graphite ball-milling. Additionally, quantitative topological characteristics of the short- and medium-range orders were estimated using RMC modeling and Voronoi analysis based on the experimental X-ray data.

## 2. Experimental details and modeling procedure

Crystalline spectrally pure graphite powder (99.99 % C) was used as the starting material. The graphite powder of ~ 6.7 g and 15 silicon nitride grinding balls were charged into a 250 ml bowl made from sintered silicon nitride with a steel casing. Graphite was milled in argon gas atmosphere in the laboratory planetary ball mill Fritsch Pulverisette P-6 at the room temperature for 0.5, 1, 2, 3.5, 10 and 20 hours, respectively. The rotation speed was 400 rpm for all experiments. The ratio of balls to sample weight was 30:1. The specific energy input was estimated accordingly to the test-object method [39] for the same conditions that were used in all our experiments and it is equal to 2.42 $Jg^{-1}s^{-1}$. The energy doses transferred to the samples are 1.07, 2.14, 4.28, 7.49, 21.46 and 42.92 kJ for 0.5, 1, 2, 3.5, 10 and 20 hrs of milling, respectively. The specific energy doses transferred to the graphite powder for the same milling times are 0.16, 0.32, 0.64, 1.12, 3.2 and 6.4 J/g, respectively. The produced powders do not contain impurities according to the chemical analysis.

X-ray diffraction studies are carried out on the standard diffractometer with filtered Cu $K_\alpha$ and monochromated $MoK_\alpha$ radiations in the Bragg-Brentano and Debye-Scherrer geometry, respectively. The experimental structure factors (SF) and the pair radial distribution functions are calculated by the procedure described in [31,40]. Transmission electron microscopy (TEM) study is performed using the high-resolution microscope JEOL JEM-2100F. Simulation of atomic configurations of graphite in initial and ball-milled state was carried out using RMCPow v.2.4 software package [41]. All configurations consist of 4000 carbon atoms placed into a configuration cell (10×10×10 unit cells in each direction). To eliminate the inevitable overlap of atoms, the minimal value of their approach was fixed in the place of intersection of the left slope of the first peak of an experimental RDF with the x-axis. So, cut-off distance between carbon atoms has been set to 1.25 Å what satisfies the recommendations of [41].

Atomic configurations of strongly disordered carbon after 3.5, 10 and 20 hrs grinding were also reduced using the RMCA v. 3.14 package [41]. Quantitative characteristics of atomic configurations obtained by different RMC software packages showed a good correspondence. In the RMCPow procedure, the model structure factor $I^C(Q)$ is calculated by the equation (1) [42-44]:

$$I^{coh}(Q) = \frac{2\pi^2}{NV_{cc}} \sum_{\tau_{cc}} \left| F(\tau_{cc}) \right|^2 \frac{R(Q-\tau_{cc})}{\tau_{cc}^2} . \quad (1)$$

Here N is the number of atoms within the configuration cell, $V_{cc}$ - its volume, $\tau_{cc}$ runs over the set of reciprocal lattice points for the configuration cell and $R(Q-\tau_{cc})$ - the experimental resolution function. The amplitude sum at a reciprocal point Q is

$$F(Q)=\sum_{j=1}^{N} f_j(Q)e^{iQr_j} \tag{2}$$

and it runs over all N atoms in the configuration cell, where $f_j$ and $r_j$ are the scattering amplitude and position of atom. The difference between the simulated $I^C(Q)$ and the experimental $I^E(Q)$ structural factors have been calculated for each iteration:

$$\chi^2=\sum_{i=1}^{m}\frac{\left[I^C(Q_i)-I^E(Q_i)\right]^2}{\sigma^2(Q_i)}, \tag{3}$$

where $\sigma^2(Q_i)$ is an experimental error.

A random displacement of one atom of the model configuration is performed during the next step of the calculations. The new amplitude sums, model structure factor and new $\chi^2$ are calculated. The process is repeated at the decreasing of $\chi^2$.

The parameters of the short-range order (the 1-st coordination number $N_1$ [45] and the bond angle value $\Theta_{ijk}$ [2]) are calculated according to the equations (4) and (5):

$$N_1=\int_0^{R_{\min}} dr4\pi r^2\rho(r)g(r), \tag{4}$$

$$\Theta_{ijk}=\cos^{-1}\left(\frac{r_{ij}^2+r_{ik}^2+r_{jk}^2}{2r_{ij}r_{ik}}\right), \tag{5}$$

where $R_{min}$ is the first minimum on a radial distribution function $g^c(r)$, $r_{ij}$, $r_{ik}$ and $r_{jk}$ are the bond lengths between the nearest atoms ij, ik and jk, respectively.

Taking into account $\Theta_{ijk}$, the bond angle distributions P(Θ) are calculated by the following equation [2]:

$$P(\Theta)=\frac{1}{\sum_{i=1}^{N}N_i(N_i-1)}\sum_{i=1}^{N}\sum_{j=1}^{N_i}\sum_{k=j+1}^{N_i}\delta(\Theta-\Theta_{ijk}), \tag{6}$$

where N is the total number of atoms, $N_i$ is the number of the neighbors for the i-th atom.

The quantitative estimation of the medium-range order was performed accordingly to S. King's criterion [14,16] by procedure [41]. It allows to represent the typical atomic configurations for disorder materials as a closed rings in the ensemble of atoms. The rings are formed due to the arrangement of atoms in polygons (in the general case of irregular shape) with the limited maximal length of the chemical bond. A ring size n of the network is determined by a number of angles in a closed polygon [15]. The statistical and geometrical features of the local atomic arrangements in the generated atomic configurations were established using Voronoi polyhedral analysis [46-48] and software package developed by V. Voloshin [49, 50].

## 3. Results

The XRD patterns for graphite in the initial and ball-milled states are shown in Fig. 1. The strong crystalline peak (002) is at the position of 26.45° on the XRD pattern of pristine graphite. It is seen that the peak intensity significantly decreases after milling. Simultaneously, the peak width increases during the grinding due to gradual disordering of the crystal structure of graphite. A broad asymmetric halo on the XRD diagrams after 3.5 hrs grinding indicates the collapse of long-range order in a structure of the ball-milled graphite, what is in a good agreement with experimental structure factors behaviour (Fig.2). Decrease of intensity of the peak in a graphite position ($s_1$=1.88 Å$^{-1}$) with simultaneous increase of that in a diamond location ($s_2$=3.05 Å$^{-1}$) as well as essential their broadening testifies the formation of carbon material with mixed graphite- and diamond-like short-range order.

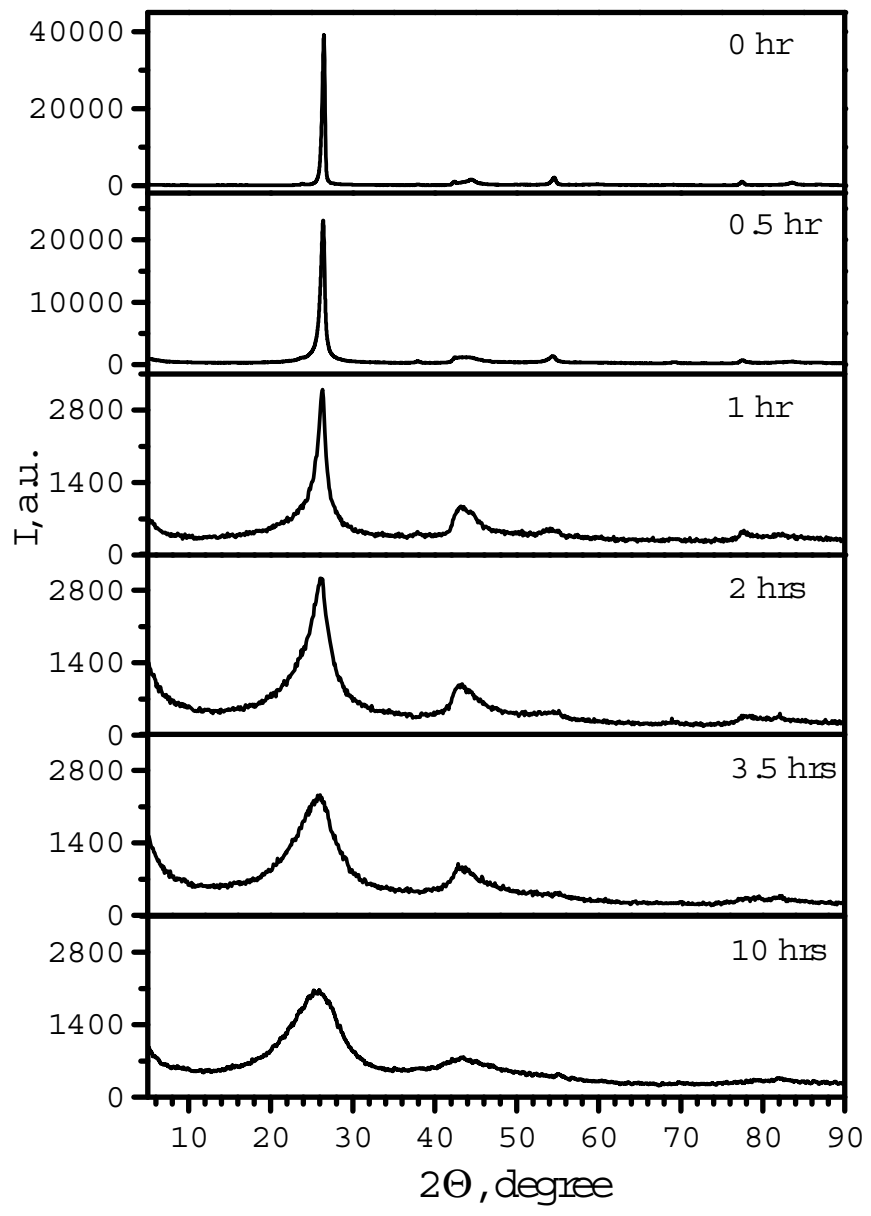

Fig. 1 – X-ray diffraction diagrams for graphite in initial state and after ball-milling.

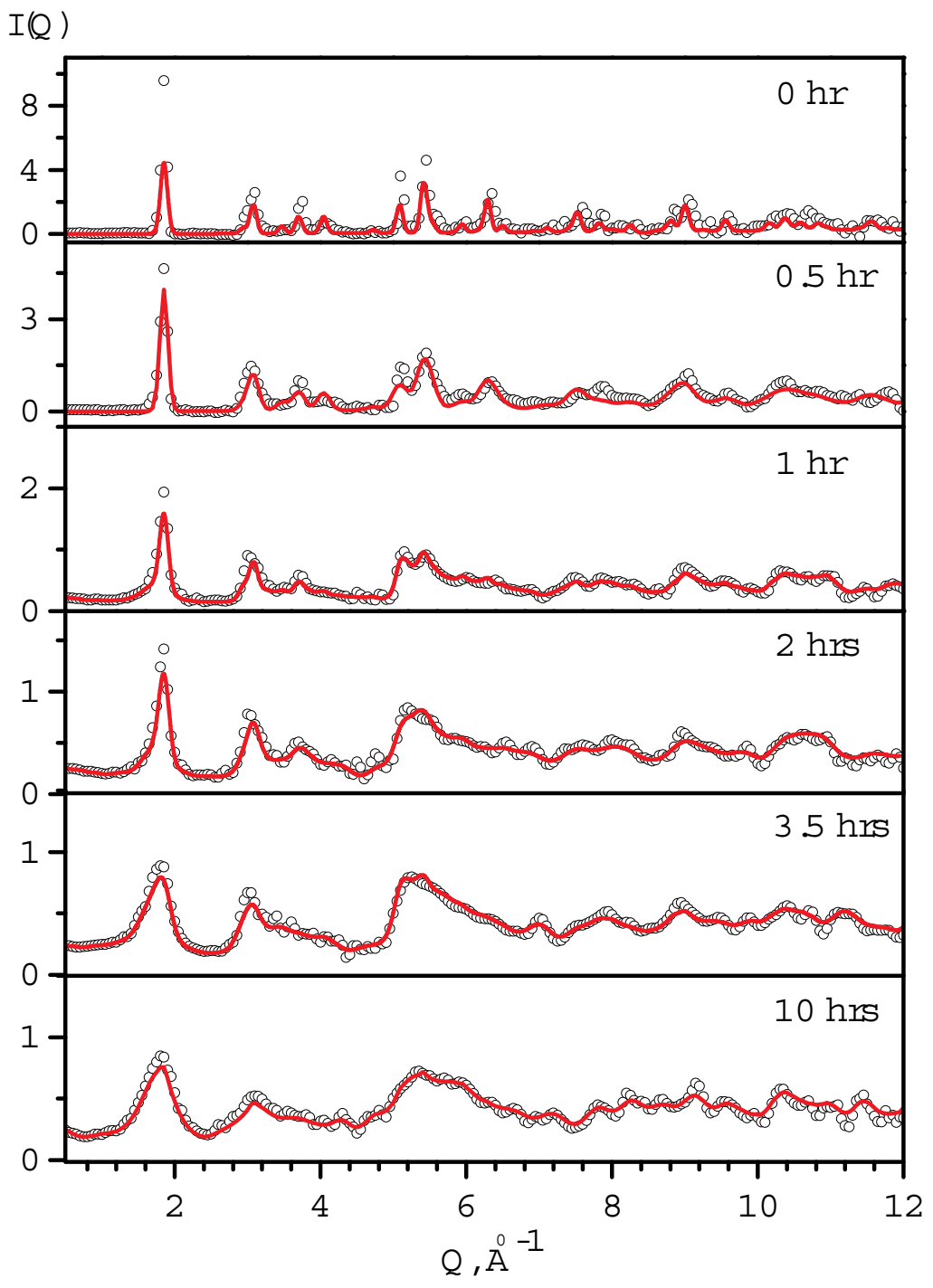

Fig. 2 – Structure factors I(Q) (open circles – experiment, solid line – modeling) of ball-milled graphite.

High resolution TEM images revealed increase disordering in graphyne layers with increase of milling time, which are still retained after 2 hrs treatment (Fig. 3a). In this case, they have bent shape. Further increase of processing time leads to the formation of disordered nanomaterials with structure typical for turbostratic carbon (Fig. 3b and c). Electron micro-diffraction patterns (Fig. 3d) are in accordance with X-ray diagrams: there are 2 broad rings confirming amorphous state in the 10 hrs ball-milled material.

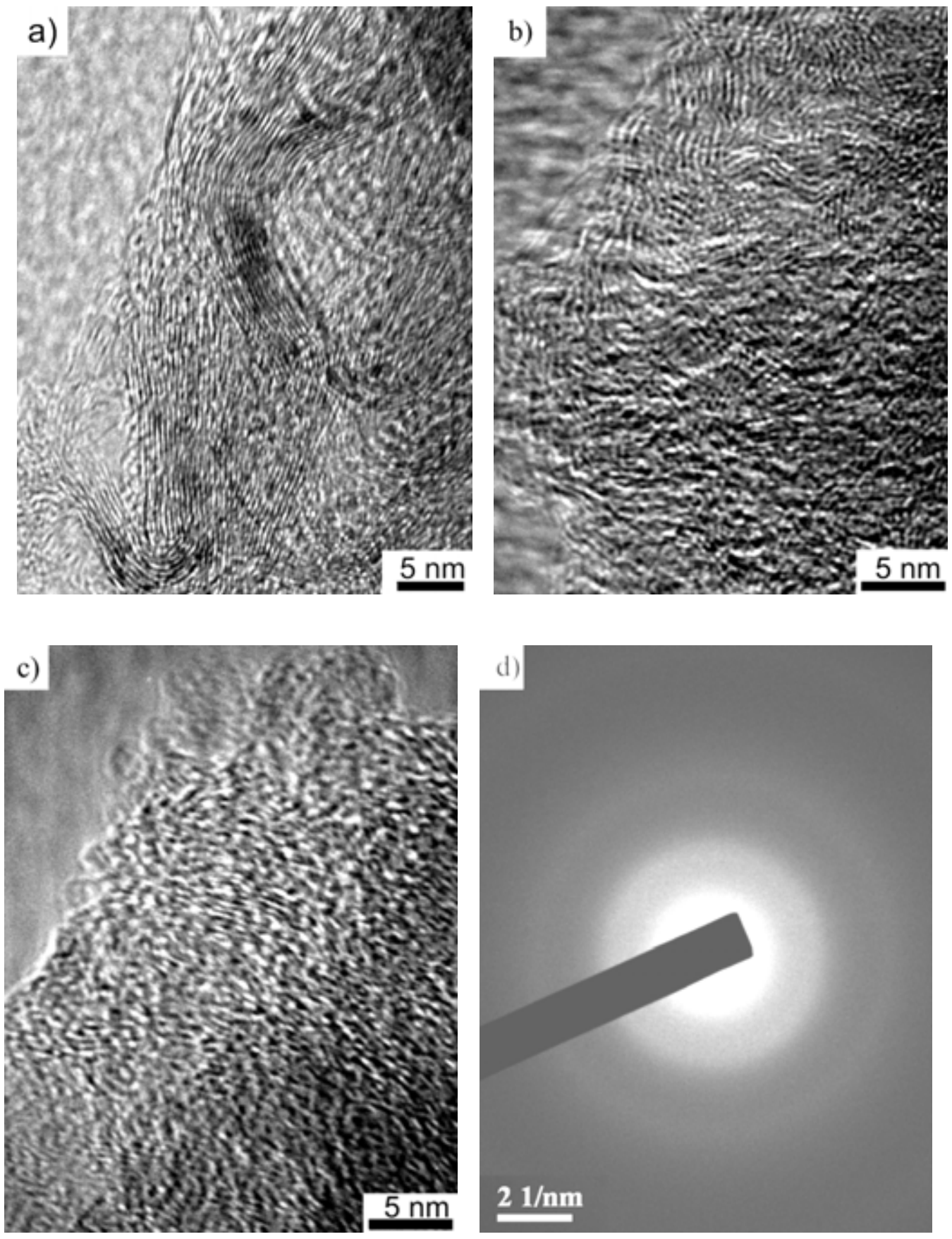

Fig. 3 – High-resolution micrographs of carbon nanomaterials obtained after ball-milling of graphite: (a) – 1 hr; (b) – 3.5 hrs; (c) – 10 hrs; (d) – electron micro-diffraction, 10 hrs.

### *3.1. Reverse Monte Carlo modeling*

Atomic configurations were generated by RMC procedure using experimental structure factors for all specimens. The agreement between the experimental and simulated structure factors indicates adequacy of the generated atomic configurations to the real structure of the samples (Fig. 3).

Spatial images of modeling structures of ball-milled graphite are shown at the Fig. 4. It is clearly seen that a continuous transformation of the perfect graphyne planes into completely disordered structure takes place during the milling process. The parameters of the short-range order (Table 1) were calculated by the equations (4) and (5). It is obvious that the experimentally determined values of the coordination numbers $N_1$ and bond angle $\Theta$ for pristine graphite are in a good agreement with the literary data [51]. Increasing the value of $N_1$ towards its diamond value ($N_1$=4) occurs in the process of grinding. Calculated distributions of the valence bond angles for the model atomic configurations as a function of graphite grinding time are shown in Fig. 5a. The intensive peak in the range of 120º is presented on the distribution for the crystalline graphite. It is characteristic for ideal hexagons which form the graphyne plane.

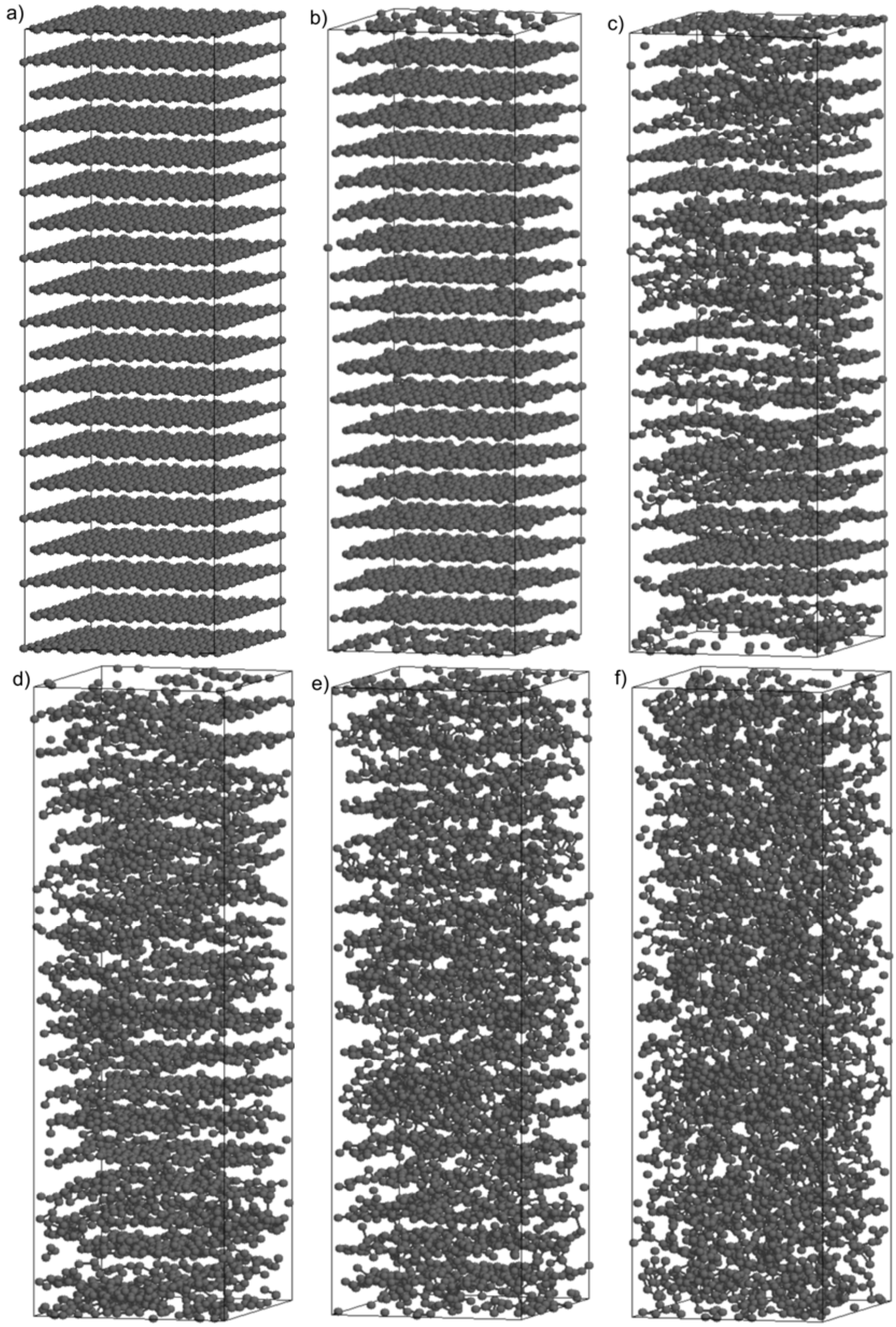

Fig. 4 – Spatial images of modeling structures of ball-milled graphite: (a) – 0 hr; (b) – 0.5 hr; (c) – 1 hr; (d) – 2 hrs; (e) – 3.5 hrs; (f) – 10 hrs.

The conspicuous peak broadening is observed after 0.5 hr grinding. It indicates slight deformation of the hexagon structure. Increase of milling time results in an appearance of a low-intensity broad peak at 57.23º, which gradually grows and shifts toward greater angles. Further increase of milling time provides both a decrease of intensity of peak at ~120° and an increase of its half-width. Such behavior of the bond angle distribution indicates about the gradual structural disorder of graphite and the respective formation of a randomly arranged close-packed structure after 10 hrs of ball-milling.

Table 1 – Structural parameters ($N_1$ – coordination number, Θ – bond angle, $K_{sph}$ – coefficient of sphericity) of the ball-milled graphite calculated for model atomic configurations.

| Ball-milling time | 0 hr | 0.5 hr | 1 hr | 2 hrs | 3.5 hrs | 10 hrs | 20 hrs |
|---|---|---|---|---|---|---|---|
| $N_1$ | 3 | 3 | 2.86 | 3.13 | 3.25 | 3.22 | 3.42 |
| Θ, degree | 119.30 | 119.85 | 57.23<br>118.23 | 55.70<br>116.59 | 58.28<br>113.87 | 58.38<br>111.19 | 57.62<br>110.9 |
| $K_{sph}$ | 0.35<br>0.37 | 0.39 | 0.39 | 0.43<br>0.48<br>0.52 | 0.45<br>0.56 | 0.47<br>0.55 | 0.48<br>0.58 |

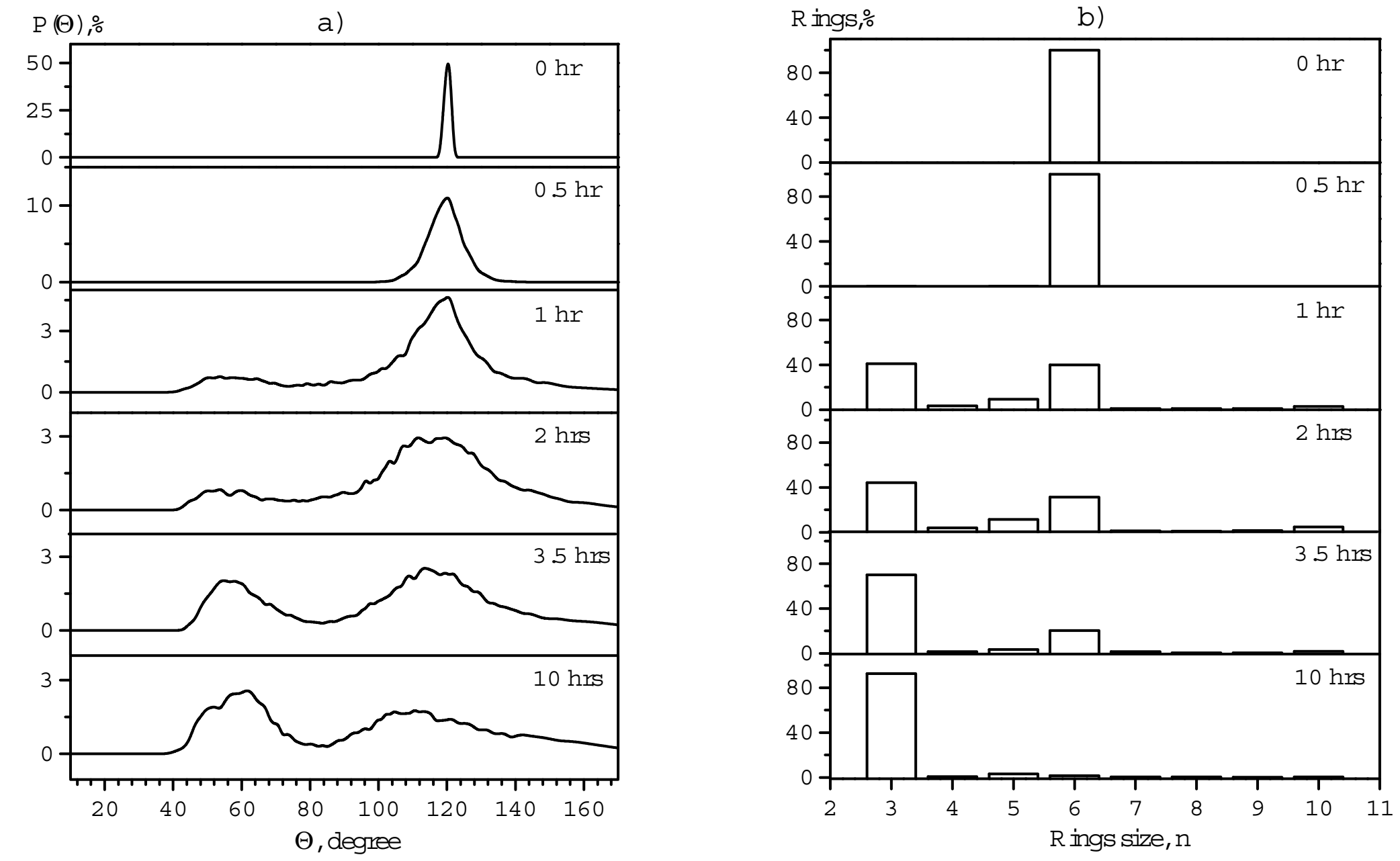

Fig. 5 – The bond angle – (a) and (b) – ring size distributions in the simulated atomic configurations for ball-milled graphite.

The features of the graphite structural transformations during grinding were analyzed on the basis of the medium-range order conception [10,23,45,52]. The statistical analysis of n-fold rings has shown that the ring size distribution (Fig. 5b) is in a good agreement with the bond angle distribution (Fig. 5a, Table 1). Therefore, 6-fold rings dominate in the pristine graphite after 0.5 hr ball-milling. Attributes of the 3-, 4-, 5- and 10-fold rings with simultaneous decreases of 6-fold ones appear in the structure of the graphite samples after 1 hr grinding. The 3-fold rings dominate in the structure of ball-milled graphite (Fig. 5b) after 3.5 hrs processing. The intensive peak at ~58° is caused by these 3-folds rings (see Fig. 5a). The 4-, 5- and 10-fold rings are presented in a very small amount. Moreover, they have a distorted shape that is due to the width and asymmetry of the second peak at ~111-119° in the bond angle distributions (Fig. 5a). The dominance of 3-fold rings indicates a randomly close-packed structure in the synthesized amorphous carbon materials and it agrees with the assumption made from the analysis of the bond angle distribution and increasing of the first coordination number.

### *3.2. Voronoi diagram analysis*

The local atomic arrangements in the simulated ensembles were examined by means of a Voronoi diagram analysis using the software package developed by V. Voloshin [49,50]. A simulated atomic space is divided into Voronoi polyhedra (VP). An individual VP is a structural unit of the amorphous state that is similar to the unit cell of the crystalline state. The basic attributes used for the statistical and geometric analyzes of disordered systems are topological (an average number of faces and edges on a polyhedron, a fraction of polyhedra with the 5-fold faces, a topological index, etc.) and metric (a volume distribution of VP, an area of the faces, a number of the neighbors at a given distance, a coefficient of sphericity, etc.) characteristics of the Voronoi polyhedra [2,46,47].

The histogram distributions of the Voronoi polyhedra versus the number of faces for the ball-milled graphite are shown in Fig. 6a. It is seen that the atomic configurations in crystalline graphite can be divided into two types of polyhedra: 50% are 11-sides and the second 50% are 15-sides. After 0.5 h grinding, Voronoi polyhedra are characterized by the wide distribution of the number of faces with predominance of 15-, 16-, 19- and 20 sides. It indicates deformation of VP typical for graphite as a result of the destruction of graphite crystal structure. Further treatment results in increasing fraction of 16 -, 17 - and 18-sided VP. The histogram after 10 hrs grinding (Fig. 6a) takes the specific for amorphous carbon nanomaterials form [36].

The 3-, 4-, 6- and 8-fold faces dominate in VP of the pristine graphite (Fig. 6b). Additional 5-, 7-, 9- and 10-fold those appears on Voronoi polyhedra after 0.5 hr grinding. The broad asymmetric distribution for the ball-milled materials indicates a wide variety of faces with different numbers of edges (corners) on each face. Voronoi polyhedra of disordered carbon are characterized by the increased amount of pentagonal faces.

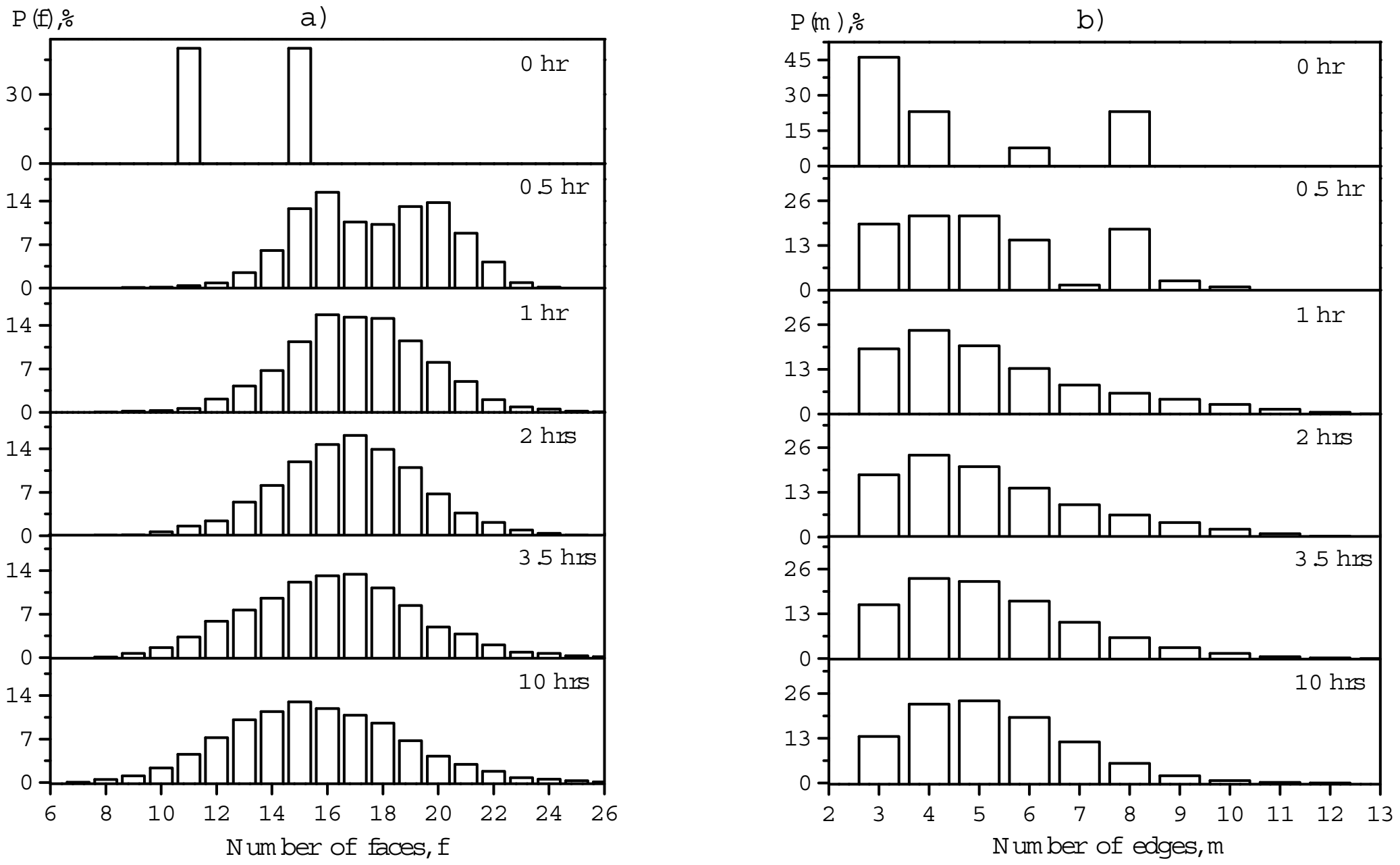

Fig. 6 – Face (a) and edge on the face (b) distributions of Voronoi polyhedra for pristine and ball-milled graphite.

The detailed analysis of the Voronoi polyhedra topology in the ball-milled graphite is carried out. An individual VP can be labeled by a topological (Voronoi) index. A topological index is defined as a set of consequtive numbers $n_3$, $n_4$, $n_5$, $n_6$, $n_7$…, where $n_i$ is the number of faces with i vertices on a polyhedron [2,46]. It is well known [36,53], that the half of VP has 11 faces: 6 - triangular, 2 - hexagonal and 3 - octagonal ones. The second half has 15 faces: 6 - triangular, 6 - tetragonal and 3 octagonal ones. However, the atomic configurations generated for the ball-milled graphite are characterized by a wide variety of the f - faces polyhedra with a prevalence of 11 - 16-sided ones (~ 44% for 2 hrs milling, ~ 51% for 3.5 hrs and ~ 58% for 10 hrs, respectively) in contrast to the crystalline structure of graphite. It is found that the polyhedra have different topologies that complicates their classification by the topological indices.

To analyze 11-16- sided VP only those found at least 7 times are selected. The selected polyhedra compose ~ 1.64%, 9.53% and 20.41% among the 11-16 sided ones formed for 2 hrs, 3.5 hrs and 10 hrs ball-milling time, respectively. Other polyhedra are found in small amounts, mainly in single numbers, i.e. every polyhedron can be specified by an individual topological index. The histogram of the topological indices of Voronoi polyhedra calculated for the 10 hrs ball-milled graphite is shown in Fig. 7.

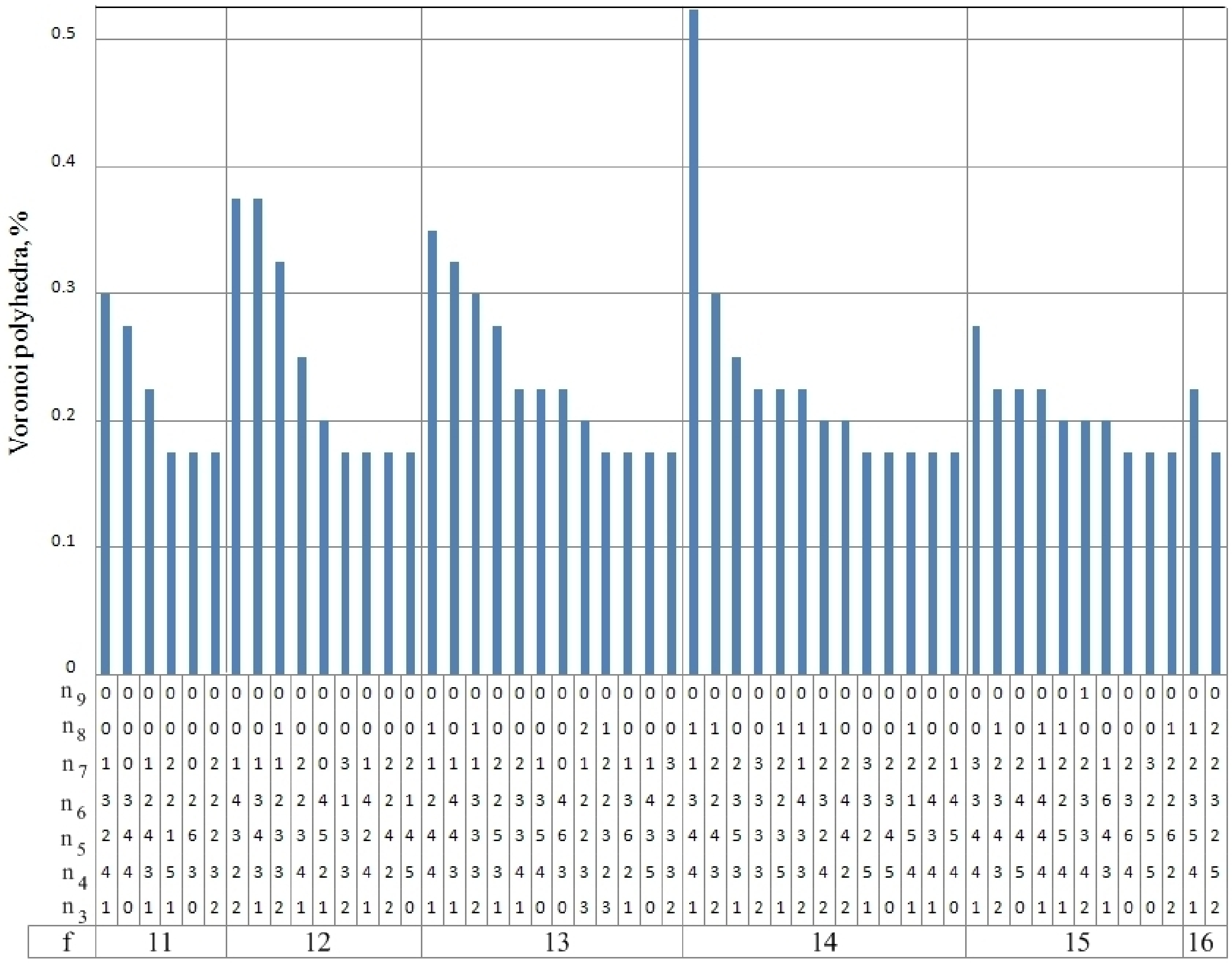

Fig. 7 – Histograms of topological (Voronoi) indices calculated for ball-milled graphite during 10 hrs; f — number of faces on Voronoi polyhedron.

It is clearly seen that a variety of topological indexes is inherented by polyhedra with the same number of the faces that indicates structures close to disordering. Moreover, 5-fold faces fitted to amorphous structures dominate in the 11 - 16-sided polyhedra. Polyhedra with 5-fold faces practically never occurs in the crystal structures. The presence of these faces provides a symmetry axis of the 5th order that is incompatible with the translational symmetry of the crystal lattice. It should be noted, that the topological indices inherented to crystalline graphite are absent in the ball-milled samples.

Of particular interest is the analysis of a sphericity coefficient $K_{sph}$ of VP ($K_{sph} = \frac{36\pi V^2}{S^3}$, where V and S are polyhedral volume and area, respectively) [46,54,55]. Graphite crystalline lattice is not a close-packed structure and it is characterized by a rather low $K_{sph}$ (Table 1). It remains steady for one hour grinding. After that time, the distribution is broadened and takes an asymmetric shape (Fig. 8). For a more detailed analysis, the profile of the plot was approximated by three Gauss-Lorentz components (Table 1). After 3.5 hrs grinding, the component of graphite disappears and the plot becomes bimodal up to 20 hrs milling. After 10 hrs grinding, a redistribution of components occurs. Thus, considered metric and topological characteristics of Voronoi polyhedra can provide a basis for quantitative characterizations of a local structure in disordered materials.

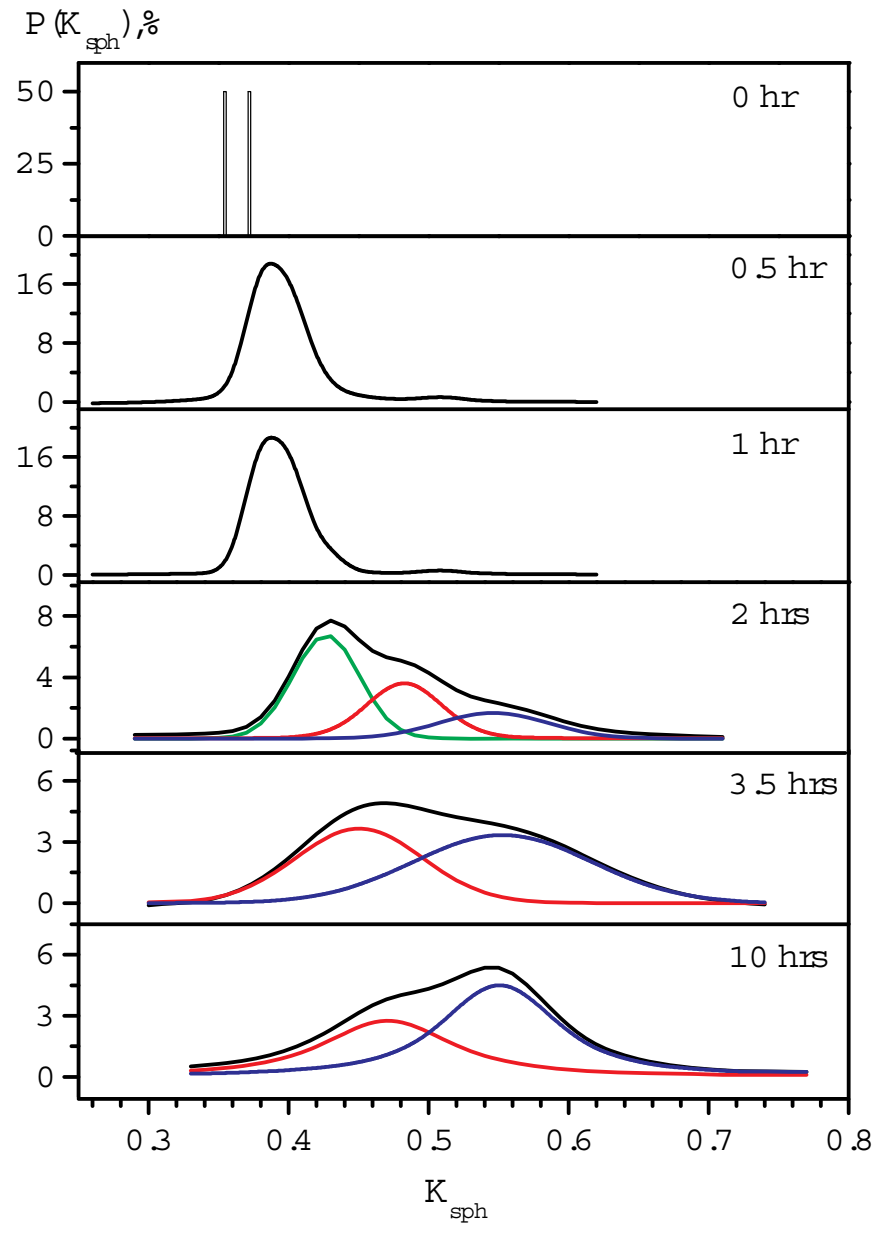

Fig. 8 – Distributions of sphericity coefficient of Voronoi polyhedra for ball-milled graphite.

## 4. Discussion

Voronoi polyhedra constructed for simulated atomic configurations of the ball-milled graphite have an extraordinary variety in their topological (Fig.6 and Fig.7) and metric (Fig.8) characteristics. First of all, the absence of predominant types of VP shows the nonexistence of a typical coordination polyhedra built on the basis of VP. The coordination polyhedra have irregular shape which contains a lot of 5-fold vertexes.

As mentioned above, there are no in the literature generally accepted quantitative characteristics to describe the topological order in amorphous materials. Based on all considered parameters describing the change of local atomic order during transition from the crystalline to amorphous state in ball-milled graphite, the behavior of VP sphericity coefficient $K_{sph}$ should be considered in the first place (Fig. 8). It is clearly seen (Table 1), that a continuous transformation of $K_{sph}$ from typical for crystalline graphite (0.35 and 0.37) to near for random system of points (0.57) takes place. In this case, the analysis of a trimodal distribution of the $K_{sph}$ for 2 hrs milled graphite has a great interest. The most intensive sub-peak is in the position for disordered graphite. There is a sufficiently large sub-peak at the position closed to the disordered tetrahedral network appropriate to distorted diamond (0.46) [36, 46]. A broad low-intensity peak appears at the position of the random system of points. The distribution of $K_{sph}$ assumes a bimodal character with a predominance of "diamond" component with increasing milling time to 3.5 hrs. Further processing for 10 hrs leads to the redistribution of intensity of components. Thus, $K_{sph}$ can be considered as the topological order parameter which allows quantitatively specifying the structure of amorphous materials: a closer value of $K_{sph}$ to the random system of points provides a more disordered structure.

Besides, the analysis of $K_{sph}$ enables a conclusion about the change in the local structure of carbon during ball-milling. So, the powder milled to 1 hr possesses the graphite-like type of short-range order. After 2 hrs grinding, in the powder structure diamond-like component appears, that increases for the material subjected to 3.5 hrs milling. The following treatment results in a decrease of this component. A performance of $K_{sph}$ curve correlates with the behavior of 2-nd peak on the bond angle distribution (Fig. 5a). It is shifted from the position characteristic of graphite (120º) to the position characteristic of diamond (109.47º). In other words, obtained structure data allow concluding that the chemical bond from $sp^2$- hybridization of the carbon atoms (graphite) changes to $sp^3$-one (diamond) in the process of graphite ball-milling. This conclusion has experimental confirmation from the analysis of Raman spectra of the studied objects and it is in accordance with [56,57]. On the other hand, thermodynamic conditions (short-time pulse pressure and temperature amounts to $\tau \sim 10^{-8}$ s, $\Delta P \sim 5$ GPa, $\Delta T \sim 1000$ K, respectively) emerging in material in a shock contact point due to grinding balls collisions in the planetary mill [58] are reasonable for the phase transformation of micro-volume graphite into diamond [59]. Moreover, an increase of milling time should lead to a destruction of the diamond-like phase and its transformation to the amorphous state what is in accordance with Fig. 8. The important role of the short-time temperature impact on the mechanism of diamond synthesis from graphite by HPHT technology [60] was firstly denoted in [61-63]. It should be pointed out that the formation of diamond-like amorphous carbon during ball-milling process and synthesis of onion-like carbon by electric discharge technique [31] occurs at the similar conditions of the short-term pulse impact of materials.

In closing, it should be remarked that additional restrictions on the relative positions of the simulated carbon atoms because of their potential interactions are used in works [64-66]. Omitting of constrains RMC usually generates configurations with most dense atomic packing and their triangle nature. For tackling this problem, the environment dependent interatomic interaction potential (EDIP) is incorporated into the package HRMC [67] as well as a constraint on the appearance of triangular rings [68]. It results in a significant change in modeling atomic coordinates up to the disappearance of the random close packing [66, 69]. But on the other hand, such approach determines properties of the generated atomic ensemble, although the true interatomic interaction potential in the produced ball-milled amorphous carbon is unknown. Therefore, the RMC method was utilized without limitations in this study.

## 5. Conclusions

It is shown that quasi-continuous structural transformation from the crystalline to amorphous state takes place in graphite during ball-milling. The parameters of a short-range order (coordination numbers, radius of the first coordination sphere and bond angle distribution) and a medium-range order (rings distribution) are established for amorphous carbon. The 1-st coordination number is increased which indicates the formation of a more close-packed structure in contrast to the pristine graphite.

The quantitative characteristics of the local atomic arrangements are determined in the amorphous structures. Statistical analysis of Voronoi polyhedra performed for generated by RMC atomic configurations has revealed that they have a wide distribution of their topological (average number of faces and edges on a polyhedron, fraction of polyhedra with 5-fold faces, topological index, etc.) and metric (volume distribution of VP, coefficient of sphericity) characteristics that are inherited by disordered materials. It is found that typical 5-fold faces for amorphous structures are predominant in the polyhedra. It should be indicated, that regular configuration polyhedra (clusters) are virtually absent in the synthesized carbon nanomaterials.

The Voronoi polyhedra for amorphous carbon are characterized by substantially higher sphericity coefficient $K_{sph}$ compared to typical for original crystalline graphite. It has bimodal distribution: one sub-peak corresponds to the structure with a disordered tetrahedral grid which is specific to diamond and the second one belongs to disordered system of points. The behavior of $K_{sph}$ and TEM investigations allows concluding that carbon nanomaterials with diamond-like short-range order with onion-like or turbostratic morphology can be produced by the ball milling.

Finally, we propose to apply the sphericity coefficient of the Voronoi polyhedra as a parameter of the topological order to quantitative estimation of disordering degree in amorphous structures.

**Acknowledgments**

Authors would like to express their heartfelt gratitude to Dr. V. Voloshin for the excess to the software for Voronoi-Delauney modeling, Dr. N. Danilenko for TEM investigation and Dr. I. Yakubtsov for helpful discussions. This work is partially supported by the joint project of NAS of Ukraine - SB RAS (#03-09-12).